\documentclass[showpacs,nofootinbib,preprintnumbers,prd,superscriptaddress,onecolumn]{revtex4-1}

\usepackage{amsmath}
\usepackage{amsfonts}
\usepackage{amssymb}
\usepackage{graphicx}
\usepackage[titletoc]{appendix}
\usepackage{color}
\usepackage{hyperref}
\usepackage{cleveref}
\usepackage[rightcaption]{sidecap}
\usepackage{subcaption}
\usepackage{comment}
\usepackage{slashed}
\usepackage{mathtools}
\usepackage{dcolumn}
\usepackage{array}
\usepackage{ctable}
\usepackage{multirow}
\usepackage{siunitx}
\usepackage{csquotes}
\usepackage{tabularx}
\usepackage{booktabs}
\usepackage{supertabular}
\usepackage{verbatim}
\usepackage{cancel}
\usepackage{graphicx}
\usepackage{dcolumn}
\usepackage{bm}
\usepackage{braket}
\DeclareMathOperator{\sinc}{sinc}

\begin{document}

\title{Entanglement generation from gravitationally produced massless vector particles during inflation}

\author{Alessio Belfiglio}
\email{alessio.belfiglio@unisi.it}
\affiliation{DSFTA, University of Siena, Via Roma 56, 53100 Siena, Italy.}

\author{Mattia Dubbini}
\email{mattia.dubbini@unicam.it}
\affiliation{Universit\`a di Camerino, Via Madonna delle Carceri, Camerino, 62032, Italy.}

\author{Orlando Luongo}
\email{orlando.luongo@unicam.it}
\affiliation{Universit\`a di Camerino, Via Madonna delle Carceri, Camerino, 62032, Italy.}
\affiliation{Department of Nanoscale Science and Engineering, University at Albany-SUNY, Albany, New York 12222, USA.}
\affiliation{Istituto Nazionale di Astrofisica (INAF), Osservatorio Astronomico di Brera, 20121 Milano, Italy.}
\affiliation{Al-Farabi Kazakh National University, Al-Farabi av. 71, 050040 Almaty, Kazakhstan.}

\date{\today}

\begin{abstract}
We study the gravitational production of spectator massless vector particles in a single-field inflationary scenario, and the related entanglement generation across the Hubble horizon. Accordingly, we consider a quasi-de Sitter background evolution, with additional metric inhomogeneities induced by the inflaton quantum fluctuations. Afterwards, we compute the corresponding production amplitude and show that it depends only on the transverse polarizations, appearing \emph{de facto} gauge-invariant, consistently with our interpretation of the vector field as the electromagnetic one. We notice that particle wavelengths turn out to be small compared to the Hubble radius, thus favoring sub-Hubble production relative to super-Hubble one. In particular, highly energetic vector particles are preferentially produced and we show that polarization effects provide a significant contribution to this behavior. Moreover, the production of nearly collinear particle pairs appears as the most probable configuration, due to the background conformal invariance of the theory and the plane-wave (massless particle-like) nature of the metric perturbation. We thus specialize our treatment to super-Hubble scales, confirming their subdominant contribution to the number density of produced particles, albeit setting a corresponding lower bound on the reheating temperature. In this scheme, we explore superhorizon entanglement between sub- and super-Hubble field modes, computing the corresponding von Neumann entropy and discussing the effects of horizon crossing on the generation of primordial entanglement.
\end{abstract}

\pacs{04.62.+v, 98.80.-k, 98.80.Cq, 03.67.bg}

\maketitle
\tableofcontents

\section{Introduction}\label{sec_intro}

Gravitational particle production (GPP) \cite{Birrell:1982ix,Ford:2021syk,RevModPhys.96.045005} associated with spectator quantum fields has attracted increasing attention in recent years, particularly in the context of cosmic inflation \cite{PhysRevD.35.2955,PhysRevD.37.3428,PhysRevD.59.023501,PhysRevD.101.083516,PhysRevD.101.123522}. When dealing with sufficiently high energy scales, GPP processes may indeed result in non-negligible particle densities, thus allowing spectator species to affect the energy budget of the universe at late times. Initial studies focused on field dynamics in unperturbed expanding spacetimes \cite{PhysRevLett.21.562, PhysRev.183.1057,PhysRevD.17.964}, typically modeled by a Friedmann-Lemaître-Robertson-Walker (FLRW) metric, and the presence of a time-dependent background has been shown to stimulate particle-antiparticle pair production from vacuum. The corresponding particle abundance can be computed via Bogoliubov transformations, which relate field modes in the far past and future, provided the background spacetime is asymptotically flat\footnote{As a less stringent condition, adiabatic expansion should be recovered in the asymptotic future. See, for example, Refs. \cite{Birrell:1982ix, PhysRevD.101.083516}.}. This mechanism has been shown to yield viable cold dark matter candidates and various quantum fields have been proposed as possible realizations, see e.g. \cite{PhysRevD.59.023501,PhysRevD.64.043503, PhysRevD.67.083514, PhysRevD.93.103520, Kannike:2016jfs, PhysRevD.96.103540, Fairbairn:2018bsw}. At the same time, GPP can affect reheating \cite{PhysRevD.64.023509, Feng:2002nb, Chun:2009yu, Bettoni:2021zhq}, baryogenesis \cite{Bambi:2006hp,Hashiba:2019mzm,Bernal:2021kaj,PhysRevD.106.075006,PhysRevD.107.063537} and even primordial black hole generation \cite{Erfani:2015rqv,Hooper:2019gtx}.

The additional presence of inhomogeneities typically enhances GPP processes from vacuum, thus leading to larger number densities \cite{Frieman:1985fr, Cespedes:1989kh, PhysRevD.45.4428}. This outcome has been confirmed in inflationary settings \cite{Belfiglio:2022qai, Belfiglio:2023rxb}, where metric perturbations associated with quantum inflaton fluctuations can provide relevant contributions to the final abundance of gravitationally produced particles. Notably, such perturbative effects become dominant when the involved quantum field exhibits a conformal symmetry in its coupling to gravity. In fact, no GPP occurs for conformally coupled fields in homogeneous and isotropic FLRW scenarios \cite{RevModPhys.96.045005}. As relevant examples, perturbative production associated with a Dirac spectator field has been studied during inflation and reheating \cite{Bassett:2001jg, Belfiglio:2025chv}, showing how a proper inclusion of spacetime perturbations would extend the allowed mass range of fermionic dark matter candidates.

GPP processes also generate quantum entanglement in the final state of the involved fields \cite{Ball:2005xa,Fuentes:2010dt, Martin-Martinez:2012chf}. Such quantum correlations are typically studied in momentum space \cite{PhysRevD.86.045014}, providing relevant information about the universe dynamics and quantum field statistics. Furthermore, a proper understanding of inflationary entanglement generation may offer valuable insights into the quantum-to-classical transition of primordial perturbations, i.e., how short wavelength quantum fluctuations are stretched by cosmic expansion, progressively losing their quantum nature \cite{Kiefer:1998qe, Burgess:2006jn,PhysRevD.102.023512,PhysRevD.102.043529}. In this picture, a protocol known as entanglement harvesting \cite{Reznik:2002fz,PhysRevD.92.064042} has been recently proposed to directly extract entanglement from quantum fields via local coupling with simpler quantum systems, usually in the form of Unruh-DeWitt detectors\footnote{A fully relativistic version of entanglement harvesting is still under investigation. In such a refined protocol, external detectors are replaced by localized quantum fields, constrained by appropriate potentials \cite{PhysRevD.109.045018}.} \cite{PhysRevD.14.870,PhysRevD.29.1047}. Although directly probing field entanglement from the background spacetime is currently unfeasible, such proposals have stimulated the search for some analogue models \cite{Barcelo:2005fc,Wilson:2011rsw, Hung:2012nc,PhysRevLett.123.180502,Vezzoli2019, Steinhauer:2021fhb}, where cosmic expansion and particle-antiparticle pair production are reproduced in laboratory settings. 

Motivated by the above outcomes, in this paper we explore GPP and entanglement generation for a spectator massless vector field within a single-field inflationary scenario. In principle, the vector field under consideration may represent any massless spectator field during inflation. In the present framework, for the sake of simplicity, we identify it with the electromagnetic field. To this end, we consider a quasi-de Sitter spacetime dynamics, where, according to the standard picture, inflaton quantum fluctuations induce inhomogeneous metric perturbations, thus breaking conformal flatness. In this scenario, we focus on scalar perturbations and study their dynamics during slow-roll. Such inhomogeneities then couple to the energy-momentum tensor of the vector field, leading to perturbative particle production processes. Particle number densities are then computed via the interaction Lagrangian associated with this coupling, while non-perturbative effects described by Bogoliubov transformations become negligible. We notice that particle wavelengths turn out to be smaller than the Hubble radius and, accordingly, particle production becomes more efficient on sub-Hubble scales. We thus point out that highly energetic particles are more likely to be produced, discussing possible polarization-induced effects. In this respect, we show that only the transverse physical polarizations contribute to the production amplitude, which is therefore gauge-invariant. In addition, the plane-wave nature of scalar metric perturbations and the background conformal invariance of the theory makes the production of collinear particles strongly favored. Afterwards, we specialize our calculations to super-Hubble contributions. Under the assumption that the sole super-Hubble production affects the universe energy density after inflation, we derive the corresponding particle number density to set a lower bound on the reheating temperature. Finally, we study the emergence of superhorizon quantum correlations during slow-roll, computing the von Neumann entropy between sub- and super-Hubble field modes and highlighting how horizon crossing affects entanglement generation during slow-roll.

The paper is outlined as follows. In Sect. \ref{sec_setup}, we introduce the inflationary scenario and the associated scalar inflaton field, deriving the dynamics of its fluctuations and the corresponding metric perturbations. In Sect. \ref{sec_production}, we add a spectator massless vector field to this picture, and study GPP effects via the Bogoliubov formalism. We then compute the probability amplitude for perturbative pair production, discussing its features and providing a physical interpretation to the predicted number density spectrum. In Sect. \ref{sec_number}, we explicitly compute the number density of produced particles, explaining why sub-Hubble production is strongly favored with respect to super-Hubble one. In Sect. \ref{sec_SuperHubble}, we specialize the calculations to super-Hubble scales, exploiting our outcomes to provide a lower bound on the reheating temperature. In Sect. \ref{sec_entanglement}, we investigate the entanglement generation between sub- and super-Hubble modes during slow-roll. Finally, in Sect. \ref{sec_conclusions}, we discuss conclusions and
perspectives of our approach.

\section{Inflationary set-up}\label{sec_setup}

We consider a scalar inflaton field $\phi$ minimally
coupled to gravity, described by the Lagrangian density
\begin{equation}
\mathcal{L}=\sqrt{-g}\left[\frac{1}{2}g^{\mu\nu}\left(\nabla_{\mu}\phi\right)\left(\nabla_{\nu}\phi\right)-\frac{1}{2}m^2\phi^2\right],
\label{lagrangianinflaton}
\end{equation}
where $m$ denotes the inflaton mass. Following standard approaches \cite{Brandenberger:2003vk}, we write 
\begin{equation}
\phi(\tau,\vec{x})=\phi_0(\tau)+\delta\phi(\tau,\vec{x}),
\label{inflatonfield}
\end{equation}
where $\tau$ denotes conformal time and $\phi_0(\tau)$ is the classical background inflaton field, with $\delta \phi$ the corresponding quantum fluctuations. 

Inflaton fluctuations induce perturbations on the background spacetime, leading to the metric tensor
\begin{equation}
g_{\mu\nu}=g_{\mu\nu}^{(0)}+\delta g_{\mu\nu}=a^2(\tau)\left(\eta_{\mu\nu}+h_{\mu\nu}\right),\quad \quad \text{with}\ \ |h_{\mu\nu}|\ll 1.
\label{metric}
\end{equation}
Focusing now on scalar degrees of freedom, we can adopt the longitudinal gauge to write the tensor $h_{\mu \nu}$ in terms of two perturbation potentials, usually labeled with $\Phi$ and $\Psi$. Moreover, in the presence of a minimally coupled scalar inflaton, we can further set  $\Phi=\Psi$. In this scenario, $h_{\mu\nu}$ takes the simple form
\begin{equation}
h_{\mu\nu}=2\Psi(\tau,\vec{x})\delta_{\mu\nu},
\label{metricperturbation}
\end{equation}
describing the (classical) scalar metric perturbations.

Once the metric tensor has been specified, first order corrections to the Einstein field equations give \cite{Riotto:2002yw}
\begin{subequations}
\begin{equation}
\Psi'+\mathcal{H}\Psi=4\pi G\phi_0'\delta\phi,
\label{Efe0i}
\end{equation}
\begin{equation}
\Psi''+2\left(\mathcal{H}-\frac{\phi_0''}{\phi_0'}\right)\Psi'-\nabla^2\Psi+2\left(\mathcal{H}'-\mathcal{H}\frac{\phi_0''}{\phi_0'}\right)\Psi=0,
\label{Efe00}
\end{equation}
\end{subequations}
where $\mathcal{H}=a^\prime/a$ and the prime denotes derivative with respect to conformal time.
As we want to study particle production during the inflationary stage, we are mainly interested in the slow-roll dynamics, thus writing the corresponding equations in this regime and introducing the slow-roll parameters $\epsilon_V$ and $\eta_V$. In particular, to evaluate $\epsilon_V$ we use the standard relation for the scalar power spectrum
\begin{equation}
P_s=\frac{1}{8\pi^2\epsilon_V}\left(\frac{H_I}{\overline{M}_{\text{Pl}}}\right)^2,
\label{powerspectrum}
\end{equation}
where $H_I$ is the Hubble parameter during inflation, which we assume approximately constant, and $\overline{M}_{\text{Pl}}$ is the reduced Planck mass. In agreement with Planck data, we consider $H_I/\overline{M}_{\text{Pl}}\simeq 2.50\times 10^{-5}$ and $P_s\simeq 2.1\times 10^{-9}$ at the pivot scale  $k_{\text{piv}}\equiv k_*\simeq 3.15\times 10^{-40}\text{ GeV}$, obtaining $\epsilon_V\simeq 3.76\times 10^{-3}$. We then take $\epsilon_V$ as constant throughout the slow-roll phase. Furthermore, within our picture, one can show that $\eta_V\simeq \epsilon_V$. Accordingly, within the slow-roll regime, we obtain
\begin{equation}
\Psi_k'+\mathcal{H}\Psi_k=\epsilon_V\mathcal{H}^2\frac{\delta\phi_k}{\phi_0'},
\label{Einsteinfieldequations0i}
\end{equation}
\begin{equation}
\Psi_k''+2\mathcal{H}\left(\eta_V-\epsilon_V\right)\Psi_k'+\left[k^2+2\mathcal{H}^2\left(\eta_V-2\epsilon_V\right)\right]\Psi_k=0,
\label{Einsteinfieldequations00}
\end{equation}
where $\Psi_k$ and $\delta\phi_k$ represent the single Fourier components of $\Psi$ and $\delta\phi$ respectively. Moreover, the background dynamics of the inflaton field simplifies to
\begin{equation}
3\mathcal{H}\phi_0'=-m^2a^2\phi_0,
\label{EOMinflatonbackground}
\end{equation}
whose solution turns out to be
\begin{equation}
\phi_0(\tau)=c|\tau|^{\kappa},
\label{inflatonbackground}
\end{equation}
with $\kappa=\eta_V/(1+v)\simeq \eta_V$. Constraining now the total number of inflationary e-foldings to 60, we readily find
\begin{equation}
N[\phi_0(\tau_f)]=\frac{\phi_0^2(\tau_f)}{4\overline{M}^2_{\text{Pl}}}-\frac{1}{2}=60\implies c\simeq 2.60\times 10^{19}\text{ GeV}.
\label{Nefolds}
\end{equation}

\subsection{The inflaton fluctuations and the metric perturbations in a de Sitter background}{\label{subsect2a}}

The equation of motion for the inflaton fluctuations can be found by perturbing the complete equation of motion for $\phi$, starting from the Lagrangian in Eq. (\ref{lagrangianinflaton}). During slow-roll, we can write
\begin{equation}
\delta\phi''_k+2\mathcal{H}\delta\phi_k'+\left[k^2+a^2m^2\right]\delta\phi_k=0,
\label{EOMinflatonfluctuation}
\end{equation}
retaining only zero-order terms in $\Psi_k$. We now select a quasi-de Sitter scale factor
\begin{equation}
a(\tau)=\frac{1}{H_I\left(2\tau_f-\tau\right)^{1+v}},
\label{scalefactor}
\end{equation}
where $v$ introduces a small correction to the de Sitter universe, turning out to be equal to the slow-roll parameter $\epsilon_V$ at first order \cite{Belfiglio:2022qai}, and $\tau \in [\tau_i,\tau_f]$, with $\tau_i$ and $\tau_f$ denoting the beginning and the end of the inflationary phase, respectively. Their numerical values can be found via the pivot scale introduced above: in particular, setting $v=0$ for simplicity\footnote{We take $\epsilon_V=v\simeq 3.76\times 10^{-3}\ll 1$, so our approximation is safe.}, and defining the shifted conformal time $\eta=2\tau_f-\tau$, we can write $\eta_*=1/k_{\text{piv}}\simeq 3.17\times 10^{39}\text{ GeV}^{-1}$, corresponding to the time when the pivot wavelengths exit the Hubble horizon. These wavelengths are constrained by Planck data to leave the horizon at least after $5$ e-folds after the beginning of inflation: setting $\ln\left(a_*/a_i\right)=10$ yields $\eta_i\simeq 6.99\times 10^{43}\text{ GeV}^{-1}$, where we name $a_*\equiv a(\eta_*)$ and $a_i\equiv a(\eta_i)$. Finally, as we consider the full inflationary stage to have a duration of $60$ e-folds, we find $\eta_f\simeq 6.12\times 10^{17}\text{ GeV}^{-1}$. The corresponding scales that cross the horizon at the beginning and at the end of the inflation are thus $k_i=1/\eta_i\simeq 1.43\times 10^{-44}\text{ GeV}$ and $k_f=1/\eta_f\simeq 1.63\times 10^{-18}\text{ GeV}$.

Eq. (\ref{EOMinflatonfluctuation}) is usually solved for the comoving variable $\delta\chi_k=a\delta\phi_k$. In particular, setting $\eta=2\tau_f-\tau$, Eq. (\ref{EOMinflatonfluctuation}) yields the well-known solution for $\delta\chi_k$
\begin{equation}
\delta\chi_k(\eta)=\sqrt{\eta}\left[c_1(k)H_{\nu}^{(1)}(k\eta)+c_2(k)H_{\nu}^{(2)}(k\eta)\right],
\label{deltachisolution}
\end{equation}
where $\nu\simeq 3/2+\epsilon_V-\eta_V$ and the functions $H_{\nu}^{(1)}$ and $H_{\nu}^{(2)}$ are Hankel functions of first and second type, respectively. The constants $c_1$ and $c_2$ are determined by selecting a suitable initial vacuum state. In particular, a common choice consists in employing the Bunch-Davies vacuum \cite{Bunch:1978yq,Danielsson:2003wb,Greene:2005wk}, which requires matching Eq. (\ref{deltachisolution}) in the asymptotic past, or equivalently for $k\eta\gg 1$, with the Minkowski solution, that is
\begin{equation}
\delta\chi_k(\tau)\xrightarrow{\tau \to -\infty}\frac{e^{-ik\tau}}{\sqrt{2k}}\implies \delta\chi_k(\eta)\xrightarrow{\eta \to +\infty}e^{-2ik\tau_f}\frac{e^{ik\eta}}{\sqrt{2k}},
\label{BunchDaviessolution}
\end{equation}
implying
\begin{equation}
\delta\chi_k(\eta)=\frac{\sqrt{\pi}}{2}
\label{deltachifinal}e^{i\left(\nu+\frac{1}{2}\right)\frac{\pi}{2}}e^{-2ik\tau_f}\sqrt{\eta}H_{\nu}^{(1)}(k\eta).
\end{equation}
Consequently, the solution for $\delta\phi_k$ can be easily obtained. In particular, a simple but faithful solution can be found by retaining only terms of order zero in the slow-roll parameters, that is considering $\nu\simeq 3/2$ and $v\simeq 0$. In so doing, indeed, we can use the well-known form of the first-type Hankel function of order $3/2$, finally yielding the following expression for $\delta\phi_k$
\begin{equation}
\delta\phi_k(\eta)=H_I\frac{e^{ik\left(\eta-2\tau_f\right)}}{\sqrt{2k^3}}\left(i+k\eta\right).
\label{deltaphisolution}
\end{equation}
Inserting the background inflaton solution from Eq. (\ref{inflatonbackground}) and the inflaton fluctuation mode of Eq. (\ref{deltaphisolution}) into Eq. (\ref{Einsteinfieldequations0i}), we find
\begin{equation}
\Psi'_k-\frac{\Psi_k}{\eta}=ik\psi_ke^{ik\eta}\left(1+\frac{i}{k\eta}\right),
\label{exactEinsteinfieldequations0i}
\end{equation}
where the derivative is now taken with respect to $\eta$, while we introduced $\psi_k=-i\gamma e^{-2ik\tau_f}/\sqrt{2k^3}$ and $\gamma=(\epsilon_V/\kappa)(H_I/c)\simeq 2.34\times 10^{-6}$. The exact solution of Eq. (\ref{exactEinsteinfieldequations0i}) turns out to be
\begin{equation}
\Psi_k(\eta)=\psi_ke^{ik\eta}+d_k\eta,
\label{scalarmetricperturbation}
\end{equation}
with $d_k$ an integration constant to be determined. In this respect, we remarkably notice that a linear term in time would not be a solution of Eq. (\ref{Einsteinfieldequations00}). Indeed, under the coordinate transformation $x^{\mu}\to x^{\mu}+\xi^{\mu}(x)$, the scalar perturbation potential transforms as $\Psi\to \Psi+\mathcal{H}\xi^0(x)$, and, accordingly, we find\footnote{We use the fact that for the scale factor in Eq. (\ref{scalefactor}), we have $\mathcal{H}(\eta)=(1+v)/\eta\sim1/\eta$.} $\Psi_k(x)\to\Psi_k(x)+\xi^0(x)/\eta$. Therefore, choosing $\xi^0=-d_k\eta^2$ would cancel the linear term $d_k\eta$, showing that the latter depends on the adopted coordinate system and cannot be interpreted as a physical mode. For these reasons, we necessarily set $d_k=0$ and Eq. (\ref{scalarmetricperturbation}) becomes
\begin{equation}
\Psi_k(x)=\psi_ke^{ik\eta}.
\label{finalsolutionPsi}
\end{equation}

\section{Vector particle production from inhomogeneities}\label{sec_production}

We now introduce a massless vector field $A_\mu$ within the above-depicted scenario, behaving as a spectator field throughout the inflationary phase and, consequently, we can make the simplest assumption that it turns out to be the electromagnetic field. 

Accordingly, the corresponding Lagrangian density reads
\begin{equation}
\mathcal{L}_{\text{vec}}=\sqrt{-g^{}}\left(-\frac{1}{4}F^{\mu\nu}F_{\mu\nu}\right),
\label{vectoractioncurve}
\end{equation}
where indices are raised and lowered through the full metric $g_{\mu\nu}$. Expanding Eq. (\ref{vectoractioncurve}) up to first order in $\delta g_{\mu\nu}$, we find
\begin{equation}
\mathcal{L}_{\text{vec}}=\mathcal{L}_{\text{vec}}^{(0)}+\frac{\partial\mathcal{L}_{\text{vec}}}{\partial g^{\mu\nu}}\bigg{|}_0\delta g^{\mu\nu}=\mathcal{L}_{\text{vec}}^{(0)}+\frac{1}{2}\sqrt{-g^{(0)}}T_{\mu\nu}^{(0)}\delta g^{\mu\nu},
\label{vectoractioncurveexpansion}
\end{equation}
where $\mathcal{L}_{\text{vec}}^{(0)}$ is the Lagrangian density in Eq. (\ref{vectoractioncurve}), computed with respect to the background metric $g_{\mu\nu}^{(0)}$, while $T_{\mu\nu}^{(0)}$ is the zero-order energy-momentum tensor associated with the vector field. This expansion naturally defines the interaction Lagrangian
\begin{equation}
\mathcal{L}_I=\frac{1}{2}\sqrt{-g^{(0)}}T^{(0)}_{\mu\nu}\delta g^{\mu\nu},
\label{interactionLagrangian}
\end{equation}
describing, \emph{de facto}, the production of vector particle pairs from the classical perturbed background. In order to quantify particle production effects, we proceed through standard quantization, introducing creation and annihilation operators $\hat{a}_{\vec{k},\lambda}^{\dag}$ and $\hat{a}_{\vec{k},\lambda}$, satisfying standard commutation relations
\begin{equation}
\left[\hat{a}_{\vec{k},\lambda},\hat{a}_{\vec{k}',\lambda'}^{\dag}\right]=\delta_{\lambda\lambda'}\delta^{(3)}\left(\vec{k}-\vec{k}'\right).
\label{commutation}
\end{equation}
The GPP mechanism is based on the fact that, in an expanding universe, the vacuum state is not uniquely defined. In particular, assuming asymptotic flatness\footnote{A less stringent condition, typically satisfied in realistic cosmological scenarios, is adiabaticity, see e.g. \cite{Birrell:1982ix,PhysRevD.101.083516}}, we define the "in" and "out" vacua by $\hat{a}_{\vec{k},\lambda}\ket{0}_{\text{in}}=\hat{b}_{\vec{k},\lambda}\ket{0}_{\text{out}}=0$, where the involved ladder operators are related by Bogoliubov transformations
\begin{equation}
\hat{b}_{\vec{k},\lambda}=\alpha_k\hat{a}_{\vec{k},\lambda}+\beta_k^*\hat{a}^{\dag}_{-\vec{k},\lambda},
\label{Bogoliubov}
\end{equation}
with $\alpha_k$ and $\beta_k$ the Bogoliubov coefficients, satisfying $|\alpha_k|^2-|\beta_k|^2=1$, $\alpha_{-k}=\alpha^*_k$ and $\beta_{-k}=\beta^*_k$.

We assume the field to be in the "in" vacuum in the asymptotic past. Accordingly, the corresponding particle state $\ket{\psi}$ satisfies $\ket{\psi\left(\eta\to-\infty\right)}\equiv\ket{0}_{\text{in}}$. Then, within the interaction picture, the first-order evolution of the initial state is driven by the $S$-matrix Dyson expansion obtained from Eq. (\ref{interactionLagrangian}), 
\begin{equation}
\hat{S}=1+\frac{i}{2}\int d^4x \sqrt{-g}\delta g^{\mu\nu}\hat{T}\left\{\hat{T}_{\mu\nu}\right\}.
\label{Smatrix}
\end{equation}
Since the interaction Lagrangian only allows particle pair production, we find\footnote{The sum over the polarizations only includes the two physical transverse modes.}
\begin{equation}
\ket{\psi}=\hat{S}\ket{0}_{\text{in}}=N\left(\ket{0}_{\text{in}}+\frac{1}{2}\sum_{r,s=1}^2\int d^3kd^3p{}_{\text{in}}\braket{k,r;p,s|\hat{S}|0}_{\text{in}}\ket{k,r;p,s}_{\text{in}}\right),
\label{particlestate}
\end{equation}
where $\vec{k}$ and $\vec{p}$ label the three-dimensional momenta, while $r$ and $s$ the polarizations.

The number density of produced particles is thus expressed by the relation
\begin{equation}
n=\frac{1}{(2\pi a)^3}\sum_{\lambda=1,2}\int d^3q\braket{\psi|\hat{b}^{\dag}_{\vec{q},\lambda}\hat{b}_{\vec{q},\lambda}|\psi}=n_0+n_1+n_2,
\label{totalnumberdensityofparticles}
\end{equation}
where $n_0$ is the standard non-perturbative contribution, $n_1$ arises from the \enquote{interference} between zero- and two-particle states, while $n_2$ is proportional to the probability of pair production associated with perturbations. 

In particular, we remark that $n_0$ and $n_1$ vanish for null Bogoliubov coefficients.  We then observe that the Lagrangian in Eq. (\ref{vectoractioncurveexpansion}) is conformally invariant, and thus it is equal to
\begin{equation}
\mathcal{L}_{\text{vec}}=-\frac{1}{4}F^{\mu\nu}F_{\mu\nu}-\frac{1}{2}T_{\mu\nu}^{(\text{flat})}h^{\mu\nu},
\label{vectoractionflat}
\end{equation}
where now $F_{\mu\nu}=\partial_{\mu}A_{\nu}-\partial_{\nu}A_{\mu}$ and $T_{\mu\nu}^{(\text{flat})}$ is the energy-momentum tensor computed with respect to the Minkowski metric. The conformal invariance of the action makes the Bogoliubov coefficients vanish \cite{RevModPhys.96.045005}, implying $n_0=n_1=0$ in our scenario. Consequently, vector particle production only arises from inhomogeneities, without background contributions associated with the unperturbed spacetime expansion. Accordingly, at second perturbative order, one finds
\begin{equation}
n=n_2=\frac{1}{(2\pi a)^3}\sum_{r,s=1}^2\int d^3k\int d^3p \left|\braket{k,r;p,s|\hat{S}|0}\right|^2,
\label{numberdensityparticles}
\end{equation}
where $a$ is the scale factor. Eq. (\ref{numberdensityparticles}) displays the number density of vector particles produced due to the gravitational coupling between the spectator vector field and metric perturbations. Since inhomogeneities break translational invariance, each pair contains particles with generic momenta $k$ and $p$. Summing over polarizations, we find
\begin{equation}
\left|\braket{k,p|\hat{S}|0}\right|^2\equiv\sum_{r,s=1}^2 \left|\braket{k,r;p,s|\hat{S}|0}\right|^2,
\label{amplitude}
\end{equation}
thus representing the total production amplitude for a particle pair with momenta $k$ and $p$. We can see that the amplitude depends only on the transverse, and so physical, polarizations and accordingly the result is consistent with the gauge invariance.

In order to explicitly compute this probability amplitude, we start again from Eq. (\ref{Smatrix}), where we remark that the scalar metric perturbation is treated as classical, while the vector field is quantized. Because of conformal invariance, vector field quantization works through the canonical, flat-space procedure. In particular, we choose to work in Coulomb gauge, obtaining
\begin{equation}
\hat{A}_{\mu}(x)=\sum_{\lambda=1}^2\int \frac{d^3q}{(2\pi)^3}\frac{1}{\sqrt{2q}}\left[\epsilon_{\mu}\left(\vec{q},\lambda\right)\hat{a}_{\vec{q},\lambda}e^{-iq_{\mu}x^{\mu}}+\epsilon^*_{\mu}\left(\vec{q},\lambda\right)\hat{a}^{\dag}_{\vec{q},\lambda}e^{iq_{\mu}x^{\mu}}\right],
\label{quantizedvectorfield}
\end{equation}
where the polarization vector is $\epsilon^{\mu}=\left(0,\vec{\epsilon}\right)$, and its spatial part $\vec{\epsilon}\left(\vec{q},\lambda\right)$ satisfies

\begin{itemize}

\item[-] orthogonality with respect to the direction of propagation of the particle, thus $\vec{q}\cdot\vec{\epsilon}\left(\vec{q},\lambda\right)=0$,

\item[-] orthonormality with respect to its complex conjugate, that is $\vec{\epsilon}\left(\vec{q},\lambda\right)\cdot\vec{\epsilon}^*\left(\vec{q},\lambda'\right)=\delta_{\lambda,\lambda'}$,

\item[-] the completeness relation
\begin{equation}
\sum_{\lambda=1}^2\epsilon_i\left(\vec{q},\lambda\right)\epsilon_j^*\left(\vec{q},\lambda\right)=\delta_{ij}-\frac{q_iq_j}{q^2}.
\label{completenessrelation}
\end{equation}

\end{itemize}

\noindent Once obtained the vector field $\hat{A}_{\mu}$, we can readily derive the corresponding energy-momentum tensor
\begin{equation}
\hat{T}_{\alpha\beta}=\frac{1}{a^2}\left\{\frac{1}{4}\eta_{\alpha\beta}\eta^{\rho\mu}\eta^{\sigma\nu}\hat{F}_{\mu\nu}\hat{F}_{\rho\sigma}-\eta^{\mu\nu}\hat{F}_{\alpha\mu}\hat{F}_{\beta\nu}\right\},
\label{energyimpulsetensor}
\end{equation}
giving
\begin{equation}
\hat{S}=1+i\int d^4x\Psi(x)\left[\frac{1}{2}\eta^{\rho\mu}\eta^{\sigma\nu}\hat{T}\left\{\hat{F}_{\mu\nu}\hat{F}_{\rho\sigma}\right\}+\eta^{\mu\nu}\delta^{\alpha\beta}\hat{T}\left\{\hat{F}_{\alpha\mu}\hat{F}_{\beta\nu}\right\}\right].
\label{Smatrixgeneral}
\end{equation}
The electromagnetic tensor is found to be
\begin{equation}
\hat{F}_{\mu\nu}=-i\int\frac{d^3q}{(2\pi)^3}\frac{1}{\sqrt{2q}}\sum_{\lambda=1}^2\left[q_{\mu}\epsilon_{\nu}\left(\vec{q},\lambda\right)-q_{\nu}\epsilon_{\mu}\left(\vec{q},\lambda\right)\right]\left[\hat{a}_{\vec{q},\lambda}e^{-iq_{\mu}x^{\mu}}-\hat{a}^{\dag}_{\vec{q},\lambda}e^{iq_{\mu}x^{\mu}}\right],
\label{electromagnetictensor}
\end{equation}
where, for simplicity, we consider real polarization vectors, namely $\vec{\epsilon}\left(\vec{q},\lambda\right)=\vec{\epsilon}^*\left(\vec{q},\lambda\right)$. Accordingly, we find
\begin{equation}
\left|\braket{k,r;p,s|\hat{S}|0}\right|^2=\left|\int d^4x\Psi(x)\frac{e^{i\left(k_{\mu}+p_{\mu}\right)x^{\mu}}}{\sqrt{kp}}\left[k_ip_j-\left(kp+\vec{k}\cdot\vec{p}\right)\delta_{ij}\right]\epsilon_i\left(\vec{p},s\right)\epsilon_j\left(\vec{k},r\right)\right|^2,
\label{singleamplitudecomputed}
\end{equation}
where $k\equiv |\vec{k}|$ and $p\equiv |\vec{p}|$. Summing over polarizations and inserting the Fourier expansion of the scalar metric perturbation, we find, see Appendix \ref{appendixA}:
\begin{equation}
\left|\braket{k,p|\hat{S}|0}\right|^2=\frac{2\left(kp+\vec{k}\cdot\vec{p}\right)^2}{kp}\left|\int_{\tau_i}^{\tau_f}d\tau\Psi_{\left|\vec{k}+\vec{p}\right|}(\tau)e^{i\left(k+p\right)\tau}\right|^2.
\label{averagedamplitudecomputed}
\end{equation}
In order to compute the integral in Eq. (\ref{averagedamplitudecomputed}), we need to use the time-dependent solution for the single Fourier component of the scalar metric perturbation, given by Eq. (\ref{finalsolutionPsi}). We then find
\begin{equation}
\left|\braket{k,p|\hat{S}|0}\right|^2=4\gamma^2\lambda^2\frac{kp\left(1+x\right)^2}{\left(k^2+p^2+2kpx\right)^{\frac{3}{2}}}\sinc^2\left[\lambda\left(k+p-\sqrt{k^2+p^2+2kpx}\right)\right],
\label{finalamplitude}
\end{equation}
where $\lambda=\left(\eta_i-\eta_f\right)/2$, $x=\cos\left(\theta\right)$ and $\theta$ is the angle between the vectors $\vec{k}$ and $\vec{p}$. 

The amplitude in Eq. (\ref{finalamplitude}) vanishes for $x=-1$, i.e., for particles produced with antiparallel momenta, further showing a narrow peak for $x=1$, corresponding to particles produced with parallel momenta. In particular:

\begin{itemize}

\item[-] It is straightforward to see that the vanishing of the amplitude for $x=-1$ is caused by the factor $kp(1+x)^2$. From a physical standpoint, such a factor arises from summing over all the possible polarizations of the produced particles. Thus, we can claim that the impossibility of producing particles with antiparallel momenta at first order in Dyson expansion can be traced back to polarization effects.

\item[-] On the other hand, the factor $kp(1+x)^2$ has a maximum for $x=1$, favoring the production of particles with parallel momenta and thus contributing to the peak observed for $x=1$.

\end{itemize}

However, this peak is mainly due to the presence of the squared cardinal sine, that would result in an exact Dirac delta if the integration time interval were infinite. In particular, the Dirac delta-like behavior of the amplitude for $x=1$ is due to 

\begin{itemize}

\item[-] the nature of the scalar metric perturbation, that behaves like an incoming massless particle with respect to the amount of energy and momentum furnished to the background,

\item[-] the conformal invariance of the free Maxwell Lagrangian and the consequent plane wave solutions for the modes of the vector field,

\item[-] the zero mass of the produced particles.

\end{itemize}

If we write perturbation modes as $\Psi_q(x)\sim e^{-i\left(q\tau-\vec{q}\cdot\vec{x}\right)}$, we can interpret a given mode as carrying energy $q$ and momentum $\vec{q}$. Labeling now particle momenta by $\vec{k}$ and $\vec{p}$, the total energy and momentum of the pair are $k+p$ and $\vec{k}+\vec{p}$, respectively. In this scenario, "energy and momentum conservation" imply
\begin{equation}
q=\begin{cases}
|\vec{q}|=|\vec{k}+\vec{p}| & \text{from momentum conservation} \\ k+p & \text{from energy conservation}
\end{cases}\implies k+p=|\vec{k}+\vec{p}|\implies x=1.
\label{energymomentumconservation}
\end{equation}
Indeed, we should not consider Eq. (\ref{energymomentumconservation}) as a real conservation law, but rather as a kinematic interpretation of the resonance condition associated with external perturbative modes. The plane-wave nature of metric perturbation can be interpreted in terms of an incoming massless particle, i.e., a picture quite helpful to  explain the high peak obtained for $x=1$.

Nevertheless, we can clearly see from Eq. (\ref{energymomentumconservation}) that, in the absence of inhomogeneities, $\vec{q}=0$, the translational invariance would not be broken and therefore $\vec{k}=-\vec{p}$. Accordingly, the only viable configuration would correspond to antiparallel momenta, when $x=-1$. However, for $x=-1$ the amplitude would vanish because of the  factor $(1+x)^2$, obtained by summing over the transverse polarizations. In the absence of inhomogeneities, a null production amplitude is thus expected. This fact confirms our initial prescription, namely exploring gravitational vector particle production in the presence of inhomogeneities only, otherwise the whole production would vanish because of the conformal invariance of the theory.

\section{The number density of produced particles}\label{sec_number}

The amplitude in Eq. (\ref{finalamplitude}) should now be integrated over all possible momenta of the involved particles. We then introduce the standard infrared and ultraviolet cutoff scales, corresponding to $k_i$ (see Sect. \ref{subsect2a}) and $k_{\text{UV}}=a(\tau_f)M_{\text{Pl}}\simeq 3.27\times 10^{-13}\text{ GeV}$, respectively. In particular, $k_i$ denotes momentum scales crossing the Hubble horizon at the inflationary onset, so we neglect field modes which were super-Hubble before inflation started \cite{PhysRevD.102.043529}. Moreover, by virtue of rotational symmetry, we are free to fix a spatial direction along one of the two momenta and name the angle among them as $\theta$. The first and simplest approach to compute the integral is to approximate the squared cardinal sine with a Dirac delta in $x=\cos(\theta)=1$, namely
\begin{equation}
\sinc^2\left[\lambda\left(k+p-\sqrt{k^2+p^2+2kpx}\right)\right]\simeq \frac{1}{\lambda}\frac{k+p}{kp}\delta(x-1).
\label{sincDiracdelta}
\end{equation}
In so doing, we obtain the following comoving number density,
\begin{equation}
na_f^3=\frac{4\gamma^2}{\pi\lambda^3}\int_{\overline{k}_i}^{\overline{k}_{\text{UV}}} d\overline{k}\int_{\overline{k}_i}^{\overline{k}_{\text{UV}}} d\overline{p} \left[\frac{4\overline{k}^2\overline{p}^2}{\left(\overline{k}+\overline{p}\right)^2}\right]\simeq 6.34\times 10^{-19}\text{ GeV}^3,
\label{numberdensitydiracdelta}
\end{equation}
where we use the rescaled momentum variables $\overline{k}=k\lambda$ and $\overline{p}=p\lambda$, while $a_f \equiv a(\tau_f) \simeq 2.68\times 10^{-32}$.

This result can be improved by considering a smoother approximation for the squared cardinal sine, in place of the Dirac delta\footnote{The need to substitute the squared cardinal sine arises primarily from numerical considerations: the sinusoidal function oscillates very rapidly, resulting in many oscillations within the integration interval.}.

\begin{figure}
\centering
\includegraphics[width=1.\linewidth]{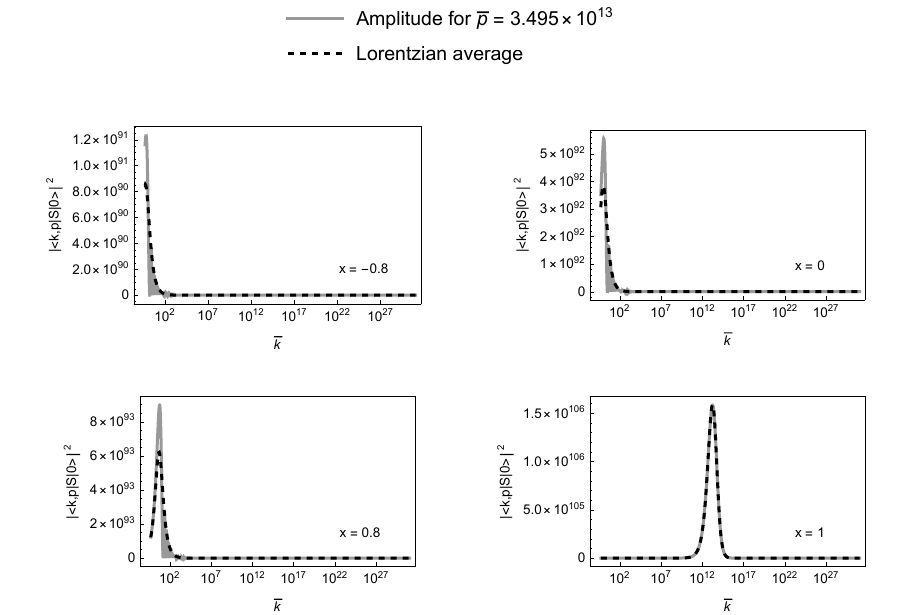}\\
\caption{Amplitude containing the oscillating squared cardinal sine in comparison to that using the average in Eq. (\ref{sincLorentzian}). The functions are shown for a fixed value of $\overline{p}$ and different values of $x$, with respect to the independent variable $\overline{k}$. A log-lin scale is used.}
\label{plotsincLorentzian}
\end{figure}

We emphasize that the above approximations are well justified in the presence of a squared cardinal sine exhibiting a sharply localized central peak. Although they do not represent exact mathematical identities, they provide accurate and computationally efficient representations for practical purposes, particularly in our numerical implementations. In this respect, and following standard techniques in optics and signal processing, the central peak of the squared cardinal sine can be reliably modeled by a Lorentzian profile \cite{Cupo2019, Robertson_2013, kauppinen1982smoothing}, yielding in our case
\begin{equation}
\sinc^2\left[\lambda\left(k+p-\sqrt{k^2+p^2+2kpx}\right)\right]\sim\frac{1}{1+\left[\lambda\left(k+p-\sqrt{k^2+p^2+2kpx}\right)\right]^2}.
\label{sincLorentzian}
\end{equation}
In Fig. \ref{plotsincLorentzian}, we show the amplitude displayed in Eq. (\ref{finalamplitude}), compared with the approximate result of Eq. (\ref{sincLorentzian}). We can see that, for $x\neq 1$, the Lorentzian curve slightly underestimates the main peak of the squared cardinal sine, despite sharing the same order of magnitude. However, the amplitude turns out to be enormously peaked around $x=1$, and in this case the two curves exactly coincide. This result shows that the Lorentzian approximation leads to the dominant peak, with increasing precision for sharper profiles. Hence, for sufficiently peaked functions, as in our case, the approximation is justified. Accordingly, the Lorentzian approximation remains accurate up to the evaluation of the amplitude integral, as the latter is almost uniquely affected by the dominant central peak. Moreover, as in $x=1$ the amplitude is almost $10-15$ orders of magnitude greater than in $x\neq 1$, we also expect the Dirac delta approximation to work rather well in estimating the number density of particles. 

In the case of a Lorentzian approximation, momentum-space integration gives
\begin{equation}
na_f^3=\frac{1}{\pi\lambda^6}\int_{\overline{k}_i}^{\overline{k}_{\text{UV}}} d\overline{k} \ \overline{k}^2\int_{\overline{k}_i}^{\overline{k}_{\text{UV}}} d\overline{p} \ \overline{p}^2\int_{-1}^1dx \left|\braket{k,p|\hat{S}|0}\right|^2\simeq 9.95\times 10^{-19}\text{ GeV}^3,
\label{numberdensitycomputed}
\end{equation}
which is consistent with the approximate result in Eq. (\ref{numberdensitydiracdelta}). This confirms our initial expectation, suggesting that the production of collinear particles is actually strongly favored.

\begin{figure}
\centering
\includegraphics[width=1.\linewidth]{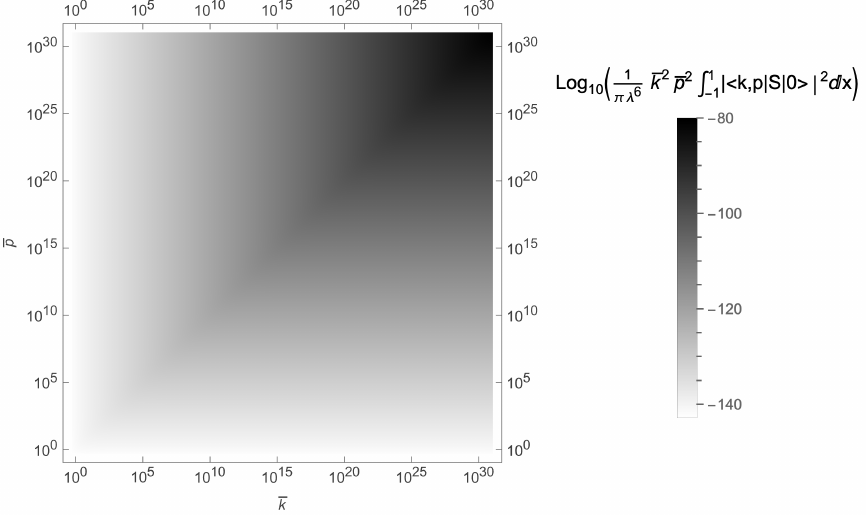}\\
\caption{Density plot showing the magnitude of the integrand function of Eq. (\ref{numberdensitycomputed}) with respect to the rescaled momenta $\overline{k}$ and $\overline{p}$. Darker regions correspond to higher values of the function, indicating momenta at which particle production is most likely.}
\label{densityplot}
\end{figure}

Moreover, it can be shown that particles are more likely to be produced with large momenta of similar magnitude, as illustrated in Fig. \ref{densityplot}. Here, we use a log-log scale; from this point forward the same scale is adopted in all figures.

In Fig. \ref{densityplot}, darker regions correspond to larger number densities. In particular, the dark region along the diagonal suggests that the production of vector particles with $\overline{k}=\overline{p}$ is favored, as expected because of the symmetry under the exchange of the two particles exhibited by the amplitude. In addition, we notice that it is typically easier to produce particles with large momenta.

\begin{figure}
\centering
\includegraphics[width=1.\linewidth]{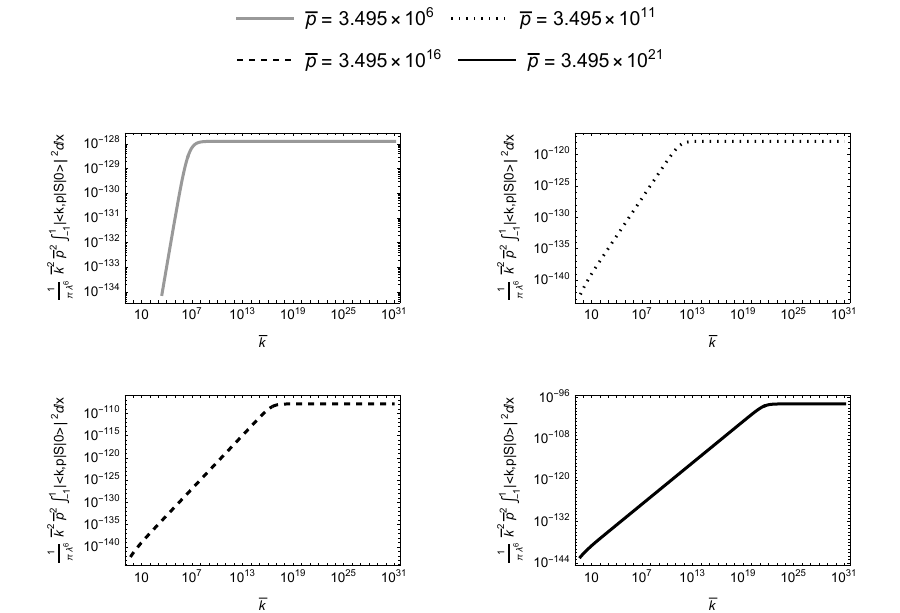}\\
\caption{Integrand function from Eq. (\ref{numberdensitycomputed}) as a function of the rescaled momentum $\overline{k}$, for different values of $\overline{p}$.}
\label{2dnumberdensity}
\end{figure}

To better understand the behavior of the integrand in Eq. (\ref{numberdensitycomputed}) as a function of particle momentum, we plot it in Fig. \ref{2dnumberdensity} with respect to $\overline{k}$, at fixed $\overline{p}$. We can see that the function grows to a maximum value, whose position changes depending on the value of $\overline{p}$. In particular, the maximum is approximately located at $\overline{k}=\overline{p}$. At larger momenta, the integrand approaches a plateau, indicating that the production of highly energetic vector particles is favored.

This fact is not obvious; previous works focusing on spin-0 and spin-$1/2$ perturbative GPP \cite{Belfiglio:2024xqt, Belfiglio:2025chv} show that the mechanism of gravitational particle production is typically more efficient at low momenta. However, in this scenario, the spin-1 nature of the spectator field and conformal invariance lead to a significant enhancement of GPP at large momenta. In particular:

\begin{itemize}

\item[-] As already discussed, summing over all polarizations results in the factor $kp(1+x)^2$. Keeping one of the two momenta fixed, this term is linear in the remaining momentum: if such a term were absent, the high-momentum behavior of the amplitude, shown in Fig. (\ref{2dnumberdensity}), would be radically modified, and it would scale inversely with the momentum. Therefore, the spin-1 nature of the field in question turns out to be fundamental in enhancing particle production at large momenta.

\item[-] The conformal invariance of the theory implies that the $\beta$ Bogoliubov coefficients, associated with spacetime expansion, vanish. As a consequence, the lower-order terms in Eq. (\ref{totalnumberdensityofparticles}), namely $n_0$ and $n_1$, vanish. The only nonzero contribution is then given by $n_2$, whose computation essentially mimics perturbation theory in flat spacetime, despite dealing here with gravitational effects. As a result, we find that particle wavelengths are typically small compared to the inflationary Hubble radius. In this sense, we can say that the related particle dynamics is 

\begin{itemize}

\item[-] local: particle production does not depend on the details of the background expansion, but rather on the specific interaction between the vector field and scalar metric perturbations;

\item[-] causal: particle production is most efficient within causally connected spacetime regions, i.e., inside the inflationary Hubble horizon.

\end{itemize}

Consequently, we expect the sub-Hubble production to be favored with respect to the super-Hubble production, as the latter is directly related to the expansion of the universe in terms of the Bogoliubov coefficients. Low-momentum modes, exiting the horizon at early times, become super-Hubble (frozen) rather soon, and thus their production in our framework is disfavored. On the contrary, high-momentum modes remain sub-Hubble for a longer period and cross the horizon only near the end of inflation. These modes, therefore, are more efficiently produced in our scenario. 

\end{itemize}

\section{Super-Hubble scales}\label{sec_SuperHubble}

We now specify the above calculations to super-Hubble modes, in order to understand how perturbative production processes are suppressed on these scales. Accordingly, we set $\Psi_k(x)\simeq \psi_k$, in lieu of the complete solution of Eq. (\ref{finalsolutionPsi}). From a physical standpoint, the perturbation becomes frozen on these scales, no longer oscillating in time. Consequently, the metric perturbation no longer admits an interpretation as an incoming massless particle, thereby losing the local and causal features of the particle dynamics, as expected for super-Hubble production. On the other hand, mathematically, we can see that Eq. (\ref{energymomentumconservation}) is no longer valid, and thus the peak shown by the amplitude in Eq. (\ref{finalamplitude}) for $x=1$ disappears. Accordingly, we are left with the smoother factor $kp(1+x)^2$, which is obtained by summing over possible polarizations, that slightly favors collinear particle production. The total particle amount will be reduced accordingly.

In particular, we obtain
\begin{equation}
\left|\braket{k,p|\hat{S}|0}\right|^2_{\text{SH}}=4\gamma^2\lambda^2\frac{kp\left(1+x\right)^2}{\left(k^2+p^2+2kpx\right)^{\frac{3}{2}}}\sinc^2\left[\lambda\left(k+p\right)\right],
\label{amplitudeSH}
\end{equation}
and we notice that the only difference with respect to the general case resides in the argument of the squared cardinal sine. In this case, the argument never vanishes, and therefore the production peak at $x=1$ is no longer present. Accordingly, we can safely average $\sin^2(x)\sim 1/2$, finding
\begin{equation}
\left|\braket{k,p|\hat{S}|0}\right|^2_{\text{SH}}=2\gamma^2\frac{\left(1+x\right)^2}{\left(k^2+p^2+2kpx\right)^{\frac{3}{2}}}\frac{kp}{\left(k+p\right)^2}.
\label{averageamplitudeSH}
\end{equation}
Again, the integral over $dx$ can be exactly performed. Here, the integration limits for particle momenta differ from those employed in Eq. (\ref{numberdensitycomputed}). In particular, we need to restrict our attention to modes that cross the Hubble horizon during the inflationary phase. Therefore, while the lower bound remains equal to $k_i$, the upper bound is set to $k_f$ (see Sect. \ref{subsect2a}).

We finally obtain the following comoving number density of produced particles
\begin{equation}
n_{\text{SH}}a_f^3=\frac{1}{\pi\lambda^6}\int_{\overline{k}_i}^{\overline{k}_f} d\overline{k} \ \overline{k}^2\int_{\overline{k}_i}^{\overline{k}_f} d\overline{p} \ \overline{p}^2\int_{-1}^1dx \left|\braket{k,p|\hat{S}|0}\right|^2_{\text{SH}}\simeq 7.74\times 10^{-67}\text{ GeV}^3.
\label{numberdensitycomputedSH}
\end{equation}
This confirms that super-Hubble production is typically disfavored in our framework.

\begin{figure}
\centering
\includegraphics[width=1.\linewidth]{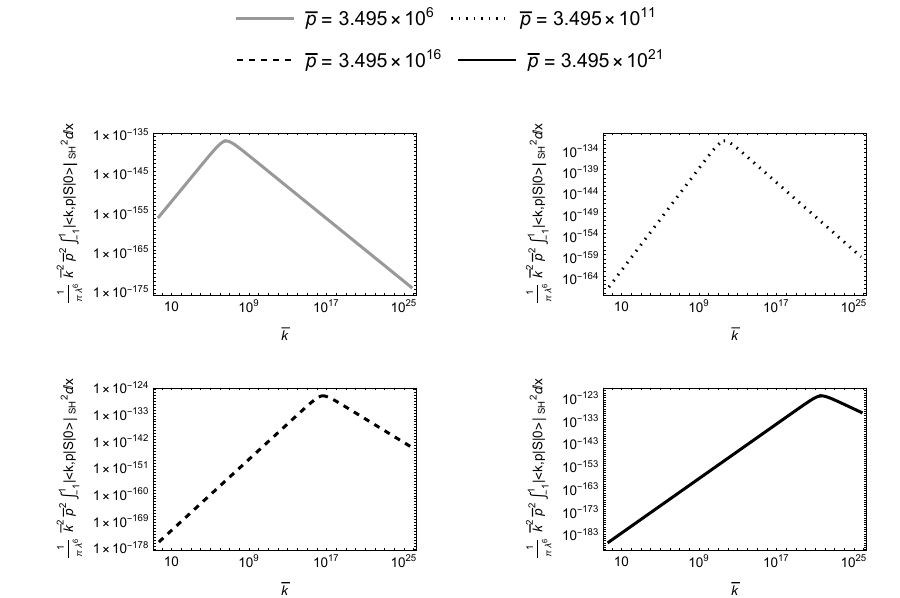}\\
\caption{Integrand function of Eq. (\ref{numberdensitycomputedSH}) with respect to the rescaled momentum $\overline{k}$ for different values of $\overline{p}$.}
\label{2dnumberdensitySH}
\end{figure}

Beyond the absence of a peak at $x=1$, the significant difference between sub- and super-Hubble production arises from the distinct approximations used for the squared cardinal sine. For sub-Hubble modes, the Dirac delta approximation yields a constant contribution, while, for super-Hubble modes, assuming $\sin^2(x)\sim 1/2$ leads to an inverse-square dependence on momentum. These differences manifest in the high-momentum behavior of the integrand functions in Eq. (\ref{numberdensitycomputed}), for general scales, and Eq. (\ref{numberdensitycomputedSH}), for super-Hubble scales. Indeed, while the former is constant (see Fig. \ref{2dnumberdensity}), the latter decreases as the inverse square of momentum, as can be seen from Fig. \ref{2dnumberdensitySH}. 

Finally, we can compute the physical number densities of produced particles corresponding to the general case and the super-Hubble limit, respectively
\begin{subequations}
\begin{equation}
n=\frac{9.95\times 10^{-19}\text{ GeV}^3}{a_f^3}=5.17\times 10^{76}\text{ GeV}^3,
\label{numberdensitycomparison}
\end{equation}
\begin{equation}
n_{\text{SH}}=\frac{7.74\times 10^{-67}\text{ GeV}^3}{a_f^3}=4.02\times 10^{28}\text{ GeV}^3,
\label{numberdensitycomparison}
\end{equation}
\end{subequations}
obtaining the following ratio: $r=n_{\text{SH}}/n\simeq 7.78\times 10^{-49}$. As previously argued, in our framework particle production is then more efficient for sub-Hubble modes.

However, only modes corresponding to super-Hubble scales are expected to freeze and can therefore contribute to the classical background energy density of the universe. On the other hand, sub-Hubble modes retain an oscillatory nature and may subsequently interact, decay or thermalize, so that their net contribution is typically suppressed. Therefore, we expect the mechanism of perturbative particle production to contribute to the classical background only through the super-Hubble component $n_{\text{SH}}$.

In particular, under the assumption that $A_{\mu}$ is the electromagnetic field, the produced massless vector particles turn out to be the photons. Accordingly, it is instructive to compare the above-derived number density with data inferred from the Cosmic Microwave Background (CMB). In particular, gravitationally produced photons may provide a lower bound on the thermal photon number density, namely
\begin{equation}
n_{\text{gravitational}}\equiv n_{\text{SH}}\le n_{\text{CMB}}\implies n_{\text{SH}}\le \frac{2\zeta(3)}{\pi^2}T^3.
\label{comparisonCMBvsgeometric}
\end{equation}
This equation allows us to set a lower bound for the reheating temperature, 
\begin{equation}
T_{\text{RH}}\ge \left(\frac{n_{\text{SH}}\pi^2}{2\zeta(3)}\right)^{\frac{1}{3}}\simeq 5.49\times 10^{9}\text{ GeV}.
\label{reheatingtemperature}
\end{equation}
The known lower and upper bounds on the reheating temperature are $\sim 10\text{ MeV}$ and $\sim 10^{16}\text{ GeV}$, respectively, set by the requirements of Big Bang Nucleosynthesis (BBN) and the inflationary energy scales \cite{RevModPhys.96.045005}. Our lower bound thus lies within the allowed range, providing a more stringent constraint in perfect agreement with the standard thermal leptogenesis mechanism, that requires $T_{\text{RH}}\gtrsim 10^9 \text{ GeV}$ to produce sufficient right-handed neutrinos for baryon asymmetry \cite{PhysRevD.108.035029}. 

In supersymmetric models, the gravitino problem typically constrains the reheating temperature to lie below $10^6-10^9 \text{ GeV}$, depending on the gravitino mass, to avoid overproducing gravitinos that destroy light elements during BBN. Nevertheless, these bounds can be relaxed in scenarios where the gravitino is the lightest supersymmetric particle. Here, the upper bound may reach $10^{10} \text{ GeV}$, and a reheating temperature $\sim 10^9 \text{ GeV}$ would be allowed \cite{Eberl:2024pxr}.

\section{Entanglement between sub- and super-Hubble modes}\label{sec_entanglement}

\begin{figure}
\centering
\includegraphics[width=0.8\linewidth]{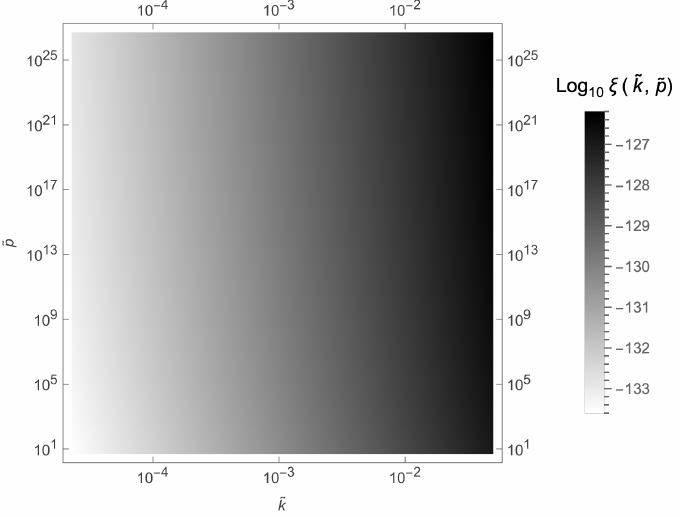}\\
\caption{Density plot showing the magnitude of the integrand function of Eq. (\ref{entanglemententropyfinal}) with respect to the rescaled momenta $\tilde{k}$ and $\tilde{p}$. Darker regions correspond to larger values of $\xi$, highlighting scales where sub- and super-Hubble modes result more entangled.}
\label{densityplotentanglement}
\end{figure}

The degrees of freedom of any interacting quantum
field theory are entangled in momentum space. Specifically, during inflation the comoving Hubble radius $r_H=1/(aH_I)$ naturally divides the total Hilbert space of states into sub-Hubble and super-Hubble subspaces, respectively. This implies that perturbative gravitational production also generates von Neumann entropy across the Hubble horizon, thus entangling sub- and super-Hubble field modes. The corresponding entanglement entropy density reads
\begin{equation}
s_{\text{ent}}=-\frac{1}{a^3}\int d^3p\int d^3k\left|\braket{k,p|\hat{S}|0}\right|^2\left[\ln\left|\braket{k,p|\hat{S}|0}\right|^2-1\right],
\label{entanglemententropy}
\end{equation}
where one integral is performed on sub- and the other on super-Hubble scales. The amplitude takes the same form as in Eq. (\ref{averagedamplitudecomputed}), where in lieu of integrating from $\tau_i$ to $\tau_f$, we select a restricted time interval $\tau\in[\tau_*,\tau_f]$, with $\tau_* > \tau_i$, in order to allow horizon crossing for a given set of modes $k < k_* = a(\tau_*) H_I$. From a mathematical standpoint, this can be simply addressed by replacing $\lambda$ with $\beta=(\eta_*-\eta_f)/2$ in Eq. (\ref{finalamplitude}), while keeping unchanged the general form of the amplitude. In the following, we select
\begin{equation}
\begin{cases}
k_i\le k \le \frac{k_*}{10} & \text{super-Hubble,} \\ 10k_*\le p \le k_{\text{UV}} & \text{sub-Hubble.}
\end{cases}
\label{integrationbounds}
\end{equation}
In order to compute the entropy density, we exploit Eq. (\ref{finalamplitude}), with the precaution of using $\beta$ in lieu of $\lambda$. For simplicity, we adopt the Dirac delta approximation for the squared cardinal sine. In so doing, we find
\begin{equation}
s_{\text{ent}}=-\frac{8\pi^2}{\beta^3a^3}\int_{\tilde{k}_i}^{\frac{\tilde{k}_*}{10}} d\tilde{k}\tilde{k}^2\int_{10\tilde{k}_*}^{\tilde{k}_{\text{UV}}} d\tilde{p}\tilde{p}^2\left(\frac{16\gamma^2}{\left(\tilde{k}+\tilde{p}\right)^2}\right)\left[\ln\left(\frac{16\gamma^2}{\left(\tilde{k}+\tilde{p}\right)^2}\right)+\ln\beta^3-1\right],
\label{entanglemententropyintermediate}
\end{equation}
where $\tilde{k}=k\beta$ and $\tilde{p}=p\beta$ are the new rescaled momentum variables, and the limits of integration are derived accordingly. The appearance of the term $\ln\beta^3$ can be traced back to the semiclassical treatment of perturbations in curved spacetime, wherein the scalar metric perturbation $\Psi$ is not quantized. In this framework, only the inflaton fluctuations and the vector field become quantum operators, while the metric perturbation is treated as a classical background. As a consequence, the normalization of the transition amplitude lacks the volume factor that would arise in a fully quantized description of all degrees of freedom. This missing contribution would exactly compensate the $\beta^3$ dependence, thus explaining the origin of the logarithmic term. For these reasons, we proceed by neglecting it in the calculation, finally obtaining the following comoving entanglement entropy density
\begin{equation}
s_{\text{ent}}a^3=-\frac{8\pi^2}{\beta^3}\int_{\tilde{k}_i}^{\frac{\tilde{k}_*}{10}} d\tilde{k}\int_{10\tilde{k}_*}^{\tilde{k}_{\text{UV}}} d\tilde{p}\left(\frac{16\gamma^2\tilde{k}^2\tilde{p}^2}{\left(\tilde{k}+\tilde{p}\right)^2}\right)\left[\ln\left(\frac{16\gamma^2}{\left(\tilde{k}+\tilde{p}\right)^2}\right)-1\right]\equiv\int_{\tilde{k}_i}^{\frac{\tilde{k}_*}{10}} d\tilde{k}\int_{10\tilde{k}_*}^{\tilde{k}_{\text{UV}}} d\tilde{p}\text{ }\xi\left(\tilde{k},\tilde{p}\right).
\label{entanglemententropyfinal}
\end{equation}
In Fig. \ref{densityplotentanglement}, we show a density plot whose regions are colored depending on the values of the integrand function in the expression for the comoving entanglement entropy density. As discussed above, $\tilde{k}$ refers to super-Hubble scales, while $\tilde{p}$ to sub-Hubble ones. The plot shows that, for fixed $\tilde{k}$, the integrand is nearly independent of $\tilde{p}$, indicating a negligible sensitivity to sub-Hubble modes. This behavior is further supported by the fact that, at fixed $\tilde{k}$, the integrand approaches an approximately constant value for large $\tilde{p}$. In contrast, for any fixed $\tilde{p}$, the integrand increases with $\tilde{k}$. Specifically, we notice that this growth is quadratic in $\tilde{k}$.

\begin{figure}
\centering
\includegraphics[width=1.\linewidth]{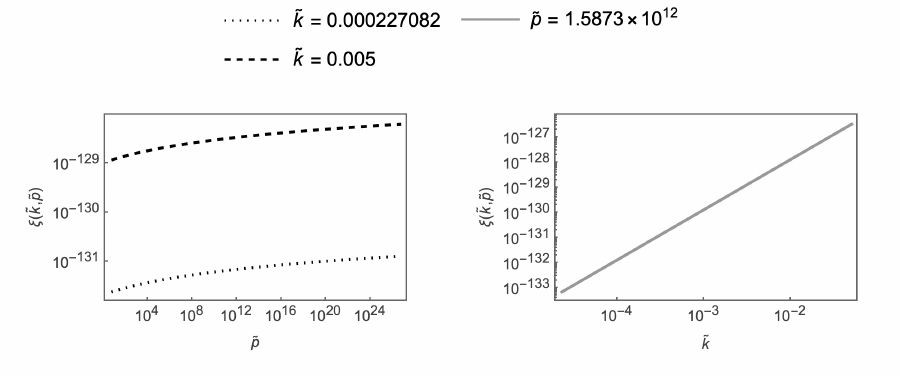}\\
\caption{Integrand function in Eq. (\ref{entanglemententropyfinal}) with respect to the rescaled momenta $\tilde{k}$ (right) and $\tilde{p}$ (left), fixing $\tilde{p}$ and $\tilde{k}$ respectively.}
\label{2dplotentanglement}
\end{figure}

In order to better visualize the above-presented scenario, in Fig. \ref{2dplotentanglement} we plot the integrand function with respect to $\tilde{k}$, fixing $\tilde{p}$, and vice versa. From the right panel, we see that, at fixed $\tilde{p}$, the entanglement entropy increases with the super-Hubble momentum $\tilde{k}$. This growth spans approximately from $10^{-133}$ to $10^{-127}$, nearly covering the full range of values observed in Fig. \ref{densityplotentanglement}. On the other hand, the left panel in Fig. \ref{2dplotentanglement} shows a very mild increase with $\tilde{p}$ at fixed $\tilde{k}$, from about $10^{-131}$ to $10^{-130}$, which is negligible compared to the overall variation of the integrand. This explains why, in Fig. \ref{densityplotentanglement}, the integrand appears effectively constant with respect to $\tilde{p}$.

The entanglement entropy computed above, typically referred to as superhorizon entanglement, is mainly generated when a given mode crosses the Hubble horizon. In fact, entanglement production is closely associated with horizon crossing. Therefore, we expect that modes remaining in sub-Hubble form throughout the inflationary phase contribute to a very small extent to the entanglement entropy, as we can check from the right plot in Fig. \ref{2dplotentanglement}. Indeed, varying $\tilde{p}$, the function $\xi(\tilde{k},\tilde{p})$ is almost constant, and its value strongly depends on the value of $\tilde{k}$. This suggests that super-Hubble modes, on the opposite, play a relevant role in the entanglement entropy. In particular, by looking at the right plot in Fig. \ref{2dplotentanglement}, we can see that the function $\xi(\tilde{k},\tilde{p})$ grows quadratically with $\tilde{k}$, for a fixed value of the sub-Hubble momentum $\tilde{p}$. Therefore, among super-Hubble modes, those with larger momenta contribute to the entanglement more than those with lower momenta, in agreement with the behavior of the particle number density in Fig. \ref{2dnumberdensity}. In particular, as vector particle production becomes more efficient at large momenta, the corresponding modes exhibit stronger entanglement.

\section{Final outlooks and perspectives}\label{sec_conclusions}

In this work, we studied the gravitational production of spectator massless vector particles and the related entanglement generation in a single-field inflationary scenario. Assuming a quasi-de Sitter background evolution, inhomogeneities were introduced in the form of scalar metric perturbations produced by the inflaton fluctuations, according to the standard picture. 

In this scenario, our spectator massless vector field is interpreted as the electromagnetic one, then described by the usual Maxwell Lagrangian. Expanding the latter up to first order in metric perturbations, it turned out to be the sum of the free Maxwell Lagrangian with respect to the conformally flat metric and an interaction term arising from metric perturbations. 

We then computed the number density of produced particles via the Bogoliubov formalism. Due to conformal invariance, non-perturbative GPP was shown to be negligible, so that particle pairs only emerge at perturbative orders, thus resembling flat-spacetime processes.

We then observed that particle wavelengths are generally smaller than the Hubble radius, so that the particle dynamics remains local and causal. Consequently, sub-Hubble production is more efficient than super-Hubble production and highly energetic particle pairs are more likely to be produced. In this respect, we proved that a relevant role is played by the spin-1 nature of the vector field, i.e., by polarization effects. In particular, we showed that only the transverse polarizations contribute to the production amplitude, consistently with gauge invariance. 

Furthermore, we showed that the produced particle pairs predominantly exhibit parallel momenta. We demonstrated that this originates from the effectively massless, particle-like behavior of the metric perturbations, taking the form of plane waves, and from the nature of the produced particles. 

Afterwards, we investigated super-Hubble production, observing that the corresponding number density is typically much smaller than the general case. We then investigated super-Hubble production, finding that the corresponding number density is typically much smaller than in the general case. We showed that this is due to the freezing of perturbations and the consequent loss of their plane-wave behavior. We used this outcome to set a lower bound to the reheating temperature, noticing that, as a first estimate, only super-Hubble (frozen) modes should contribute, in principle, to the background energy density after inflation, while sub-Hubble modes participate in the microphysical processes taking place during reheating, which inevitably affect the corresponding number densities. 

Finally, we studied the entanglement generation between sub- and super-Hubble modes. The entropy computation exhibits an extra factor with energy dimension $-3$ in the amplitude, arising from the semiclassical treatment in which metric perturbations are not quantized. In a fully quantum framework, this contribution would be canceled by a corresponding volume factor. We therefore removed it, obtaining a consistent expression for the von Neumann entropy. In this respect, we showed that larger contributions arise from super-Hubble modes, the latter crossing the horizon earlier and thus producing more entanglement.

Future works could aim to extend perturbative GPP to Proca-like fields, generalizing to vector fields with mass.  Firstly, the mass term would break conformal invariance, implying nonzero Bogoliubov coefficients. Consequently, super-Hubble production would be amplified and it would be interesting to determine its novel contribution. In future work, we also plan to extend entanglement calculations beyond the slow-roll regime, addressing thermalization effects and the resulting decoherence in various reheating scenarios. In addition, we intend to explore the role played by alternative vector fields, different from pure electromagnetism, focusing, for example, on axion Lagrangian and its GPP contribution and its impact in the current understanding of dark matter.

\section*{Acknowledgments}
AB and OL are grateful to Carlo Baccigalupi and Roberto Franzosi for interesting discussions. MD and OL acknowledge Aniello Quaranta and S. Mahesh Chandran for  debates on entanglement in quantum field theory. Finally, the authors express gratitude to Tommaso Mengoni for preliminary discussions on particle production with vector fields.

\bibliographystyle{unsrt}
\bibliography{biblio}

\appendix

\section{Calculation of the production amplitude}\label{appendixA}

Starting from the $S$-matrix in Eq. (\ref{Smatrixgeneral}), the production amplitude for two vector particles of given momenta and polarizations turns out to be
\begin{equation}
\left|\braket{k,r;p,s|\hat{S}|0}\right|^2=\left|\int d^4x\Psi(x)\left\{\frac{1}{2}\eta^{\rho\mu}\eta^{\sigma\nu}\braket{k,r;p,s|\hat{T}\left\{\hat{F}_{\mu\nu}\hat{F}_{\rho\sigma}\right\}|0}+\eta^{\mu\nu}\delta^{\alpha\beta}\braket{k,r;p,s|\hat{T}\left\{\hat{F}_{\alpha\mu}\hat{F}_{\beta\nu}\right\}|0}\right\}\right|^2.
\label{singleamplitudeappendix}
\end{equation}
In order to further simplify the expression above, we use the electromagnetic tensor of Eq. (\ref{electromagnetictensor}) and remarkably compute the following braket
\begin{equation}
\begin{split}
&\braket{k,r;p,s|\hat{T}\left\{\hat{F}_{\mu\nu}\hat{F}_{\alpha\beta}\right\}|0}=\braket{0|\hat{T}\left\{\hat{a}_{\vec{k},r}\hat{a}_{\vec{p},s}\hat{F}_{\mu\nu}\hat{F}_{\alpha\beta}\right\}|0}=\\&=-\int \frac{d^3q}{(2\pi)^3}\frac{1}{\sqrt{2q}}\int \frac{d^3h}{(2\pi)^3}\frac{1}{\sqrt{2h}}\sum_{\lambda_1,\lambda_2=1}^2\left[q_{\mu}\epsilon_{\nu}(\vec{q},\lambda_1)-q_{\nu}\epsilon_{\mu}(\vec{q},\lambda_1)\right]\times \\&\times \left[h_{\alpha}\epsilon_{\beta}(\vec{h},\lambda_2)-h_{\beta}\epsilon_{\alpha}(\vec{h},\lambda_2)\right]e^{i(q_{\mu}+h_{\mu})x^{\mu}}\braket{0|\hat{T}\left\{\hat{a}_{\vec{k},r}\hat{a}_{\vec{p},s}\hat{a}^{\dag}_{\vec{q},\lambda_1}\hat{a}^{\dag}_{\vec{h},\lambda_2}\right\}|0}=\\&=-\int \frac{d^3q}{(2\pi)^3}\frac{1}{\sqrt{2q}}\int \frac{d^3h}{(2\pi)^3}\frac{1}{\sqrt{2h}}\sum_{\lambda_1,\lambda_2=1}^2\left[q_{\mu}\epsilon_{\nu}(\vec{q},\lambda_1)-q_{\nu}\epsilon_{\mu}(\vec{q},\lambda_1)\right]\left[h_{\alpha}\epsilon_{\beta}(\vec{h},\lambda_2)-h_{\beta}\epsilon_{\alpha}(\vec{h},\lambda_2)\right]e^{i(q_{\mu}+h_{\mu})x^{\mu}}\times \\&\times \bigg{[}(2\pi)^3\delta^{(3)}(\vec{k}-\vec{q})\delta_{r\lambda_1}(2\pi)^3\delta^{(3)}(\vec{p}-\vec{h})\delta_{s\lambda_2}+(2\pi)^3\delta^{(3)}(\vec{k}-\vec{h})\delta_{r\lambda_2}(2\pi)^3\delta^{(3)}(\vec{p}-\vec{q})\delta_{s\lambda_1}\bigg{]}=\\&=-\frac{e^{i(k_{\mu}+p_{\mu})x^{\mu}}}{2\sqrt{kp}}\bigg{\{}\left[k_{\mu}\epsilon_{\nu}(\vec{k},r)-k_{\nu}\epsilon_{\mu}(\vec{k},r)\right] \left[p_{\alpha}\epsilon_{\beta}(\vec{p},s)-p_{\beta}\epsilon_{\alpha}(\vec{p},s)\right]+\left[p_{\mu}\epsilon_{\nu}(\vec{p},s)-p_{\nu}\epsilon_{\mu}(\vec{p},s)\right] \left[k_{\alpha}\epsilon_{\beta}(\vec{k},r)-k_{\beta}\epsilon_{\alpha}(\vec{k},r)\right]\bigg{\}}.
\end{split}
\label{braketuseful}
\end{equation}
Using Eq. (\ref{braketuseful}), the amplitude in Eq. (\ref{singleamplitudeappendix}) becomes
\begin{equation}
\begin{split}
&\left|\braket{k,r;p,s|\hat{S}|0}\right|^2=\bigg{|}\int d^4x\Psi(x)\frac{e^{i(k_{\mu}+p_{\mu})x^{\mu}}}{2\sqrt{kp}}\bigg{\{}\frac{1}{2}\eta^{\rho\mu}\eta^{\sigma\nu}\bigg{[}\left(k_{\mu}\epsilon_{\nu}(\vec{k},r)-k_{\nu}\epsilon_{\mu}(\vec{k},r)\right) \left(p_{\rho}\epsilon_{\sigma}(\vec{p},s)-p_{\sigma}\epsilon_{\rho}(\vec{p},s)\right)+\\&+\left(p_{\mu}\epsilon_{\nu}(\vec{p},s)-p_{\nu}\epsilon_{\mu}(\vec{p},s)\right) \left(k_{\rho}\epsilon_{\sigma}(\vec{k},r)-k_{\sigma}\epsilon_{\rho}(\vec{k},r)\right)\bigg{]}+\eta^{\mu\nu}\delta^{\alpha\beta}\bigg{[}\left(k_{\alpha}\epsilon_{\mu}(\vec{k},r)-k_{\mu}\epsilon_{\alpha}(\vec{k},r)\right) \left(p_{\beta}\epsilon_{\nu}(\vec{p},s)-p_{\nu}\epsilon_{\beta}(\vec{p},s)\right)+\\&+\left(p_{\alpha}\epsilon_{\mu}(\vec{p},s)-p_{\mu}\epsilon_{\alpha}(\vec{p},s)\right) \left(k_{\beta}\epsilon_{\nu}(\vec{k},r)-k_{\nu}\epsilon_{\beta}(\vec{k},r)\right)\bigg{]}\bigg{\}}\bigg{|}^2=\\&=\bigg{|}\int d^4x\Psi(x)\frac{e^{i(k_{\mu}+p_{\mu})x^{\mu}}}{2\sqrt{kp}} \bigg{\{}\frac{1}{2}\bigg{[}4k^{\nu}p_{\nu}\epsilon^{\mu}(\vec{k},r)\epsilon_{\mu}(\vec{p},s)-4k^{\mu}\epsilon_{\mu}(\vec{p},s)p^{\nu}\epsilon_{\nu}(\vec{k},r)\bigg{]}+\\&+\bigg{[}2\delta^{\alpha\beta}k_{\alpha}p_{\beta}\epsilon^{\mu}(\vec{k},r)\epsilon_{\mu}(\vec{p},s)-2\delta^{\alpha\beta}k_{\alpha}\epsilon_{\beta}(\vec{p},s)p^{\mu}\epsilon_{\mu}(\vec{k},r)-2\delta^{\alpha\beta}p_{\beta}\epsilon_{\alpha}(\vec{k},r)k^{\mu}\epsilon_{\mu}(\vec{p},s)+2\delta^{\alpha\beta}\epsilon_{\alpha}(\vec{k},r)\epsilon_{\beta}(\vec{p},s)k^{\nu}p_{\nu}\bigg{]}\bigg{\}}\bigg{|}^2=\\&=\bigg{|}\int d^4x\Psi(x)\frac{e^{i(k_{\mu}+p_{\mu})x^{\mu}}}{\sqrt{kp}}\bigg{\{}k^{\nu}p_{\nu}\epsilon^{\mu}(\vec{k},r)\epsilon_{\mu}(\vec{p},s)-k^{\mu}\epsilon_{\mu}(\vec{p},s)p^{\nu}\epsilon_{\nu}(\vec{k},r)+\\&+\delta^{\alpha\beta}k_{\alpha}p_{\beta}\epsilon^{\mu}(\vec{k},r)\epsilon_{\mu}(\vec{p},s)+\delta^{\alpha\beta}\epsilon_{\alpha}(\vec{k},r)\epsilon_{\beta}(\vec{p},s)k^{\nu}p_{\nu}-\delta^{\alpha\beta}k_{\alpha}\epsilon_{\beta}(\vec{p},s)p^{\mu}\epsilon_{\mu}(\vec{k},r)-\delta^{\alpha\beta}p_{\beta}\epsilon_{\alpha}(\vec{k},r)k^{\mu}\epsilon_{\mu}(\vec{p},s)\bigg{\}}\bigg{|}^2=\\&=\bigg{|}\int d^4x\Psi(x)\frac{e^{i(k_{\mu}+p_{\mu})x^{\mu}}}{\sqrt{kp}}\left[k_ip_j-(kp+\vec{k}\cdot\vec{p})\delta_{ij}\right]\epsilon_i(\vec{p},s)\epsilon_j(\vec{k},r)\bigg{|}^2,
\end{split}
\label{singleamplitudeappendix2}
\end{equation}
that is Eq. (\ref{singleamplitudecomputed}). When we sum over the polarizations and compute the modulus square explicitly, we obtain
\begin{equation}
\begin{split}
&\left|\braket{k,p|\hat{S}|0}\right|^2=\left|\int d^4x\Psi(x)\frac{e^{i(k_{\mu}+p_{\mu})x^{\mu}}}{2\sqrt{kp}}\right|^2\left[k_ip_j-(kp+\vec{k}\cdot\vec{p})\delta_{ij}\right]\left[k_lp_m-(kp+\vec{k}\cdot\vec{p})\delta_{lm}\right]\times \\&\times\sum_{r=1}^2\epsilon_j(\vec{k},r)\epsilon_m(\vec{k},r)\sum_{s=1}^2\epsilon_i(\vec{p},s)\epsilon_l(\vec{p},s)=\\&=\left|\int d^4x\Psi(x)\frac{e^{i(k_{\mu}+p_{\mu})x^{\mu}}}{2\sqrt{kp}}\right|^2\left(\delta_{jm}-\frac{k_jk_m}{k^2}\right)\left(\delta_{il}-\frac{p_ip_l}{p^2}\right)\left[k_ip_j-(kp+\vec{k}\cdot\vec{p})\delta_{ij}\right]\left[k_lp_m-(kp+\vec{k}\cdot\vec{p})\delta_{lm}\right]=\\&=2\left[kp+\vec{k}\cdot\vec{p}\right]^2\left|\int d^4x\Psi(x)\frac{e^{i(k_{\mu}+p_{\mu})x^{\mu}}}{2\sqrt{kp}}\right|^2.
\end{split}
\label{sumoverpolarizations}
\end{equation}
Finally, considering the Fourier expansion of the scalar metric perturbation, we find
\begin{equation}
\begin{split}
\left|\braket{k,p|\hat{S}|0}\right|^2&=\frac{\left[kp+\vec{k}\cdot\vec{p}\right]^2}{2kp}\int d^3q\int d^3xe^{i\left[\vec{q}-(\vec{k}+\vec{p})\right]\cdot\vec{x}}\int_{\tau_i}^{\tau_f} d\tau_x\Psi_q(\tau_x)e^{i(k+p)\tau_x}=\\&=\frac{\left[kp+\vec{k}\cdot\vec{p}\right]^2}{2kp}\left|\int_{\tau_i}^{\tau_f}d\tau\Psi_{|\vec{k}+\vec{p}|}(\tau)e^{i(k+p)\tau}\right|^2,
\end{split}
\label{amplitudeappendix}
\end{equation}
that gives Eq. (\ref{averagedamplitudecomputed}).

\end{document}